\documentclass[10pt,journal]{IEEEtran}
\IEEEoverridecommandlockouts

\usepackage{enumitem}
\usepackage[justification=centering]{caption}
\usepackage[utf8]{inputenc}
\usepackage{longtable}
\usepackage{lineno}
\usepackage[T1]{fontenc}
\usepackage{setspace}
\usepackage{supertabular}
\usepackage{tabularx,ragged2e} 
\usepackage{hyperref}
\usepackage{cleveref}
\usepackage{xspace}

\newcommand{\mainPatterns}{16\xspace}

\newcommand{\designBestPractice}{8\xspace}
\newcommand{\codingBestPractice}{18\xspace}
\newcommand{\testingBestPractice}{6\xspace}
\newcommand{\totBestPractice}{32\xspace}

\newcommand{\otherPatterns}{6\xspace}

\newcommand{\totUsefulItems}{48\xspace}
\newcommand{\totItems}{54\xspace}

\begin{document}

\title{Security checklists for Ethereum smart contract development: patterns and best practices}

\author{
\IEEEauthorblockN{Lodovica Marchesi\IEEEauthorrefmark{1}\IEEEauthorrefmark{2},
Michele Marchesi\IEEEauthorrefmark{1}, Livio Pompianu\IEEEauthorrefmark{1} and 
Roberto Tonelli\IEEEauthorrefmark{1} \\
\IEEEauthorblockA{\IEEEauthorrefmark{1}\textit{Dept. of Mathematics and Computer Science} \\
\textit{University of Cagliari}\\
Cagliari, Italy \\
Email: \{ lodovica.marchesi, marchesi, pompianu.livio \}@unica.it,  roberto.tonelli@dsf.unica.it}
\IEEEauthorrefmark{2}\textit{corresponding author}
}}

\maketitle

\begin{abstract}
In recent years Smart Contracts and DApps are becoming increasingly important and widespread thanks to the properties of blockchain technology. 
In  most  cases  DApps  are  business  critical,  and  very  strict security  requirements  should  be  assured. Developing safe and reliable Smart Contracts, however, is not a trivial task. Several researchers have studied the security issues, however none of these provide a simple and intuitive tool to overcome these problems.
In this paper we collected a list of security patterns for DApps. 
Moreover, based on these patterns, we provide the reader  with  security  assessment  checklists  that  can  be  easily used  for  the  development  of  SCs.  We cover the phases of design, coding, and testing and deployment of the software lifecycle. In this way, we allow developers to easily verify if they applied all the relevant security patterns to their smart contracts. 
We focus all the analysis on the most popular Ethereum blockchain, and on the Solidity language.
\end{abstract}

\section{Introduction}
The concept of Smart Contract (SC), first introduced in 1997 by Nick Szabo~\cite{szabo1997}, has recently gained more and more attention thanks to the introduction of public blockchains, such as Bitcoin~\cite{nakamoto2008} and Ethereum~\cite{buterin2014next}.
These are the ideal environment to run SCs, due to their transparency and immutability.
A smart contract is a piece of executable code, running on all the nodes of a blockchain, guaranteed not by a central authority, but by cryptography and by blockchain technology.
Its execution has the strong constraint that all outputs and state changes resulting from execution must be the same in all nodes.
In the blockchain scenario, smart contracts (SCs) are created and executed by specific kinds of transactions, and can be used to transfer cryptocurrencies, activate new contracts, record information and also interact to external systems.
%

Nowadays, Ethereum is the most used blockchain that supports SCs, with more than 2M deployed SCs and about 1M transactions per day (August 2020)~\cite{etherscan}.
%
In Ethereum, SCs implement a Turing-complete behavior, allowing to perform deterministic complex computations. They are written in highly-specialized programming languages, among which the most popular is called Solidity~\cite{dannen2017introducing}.

Through SCs it is possible to develop a wide range of applications in both financial and non-financial sectors, these are typically called DApps (decentralized applications).
Many DApps deal with direct digital currency or token usage, that is with entities that have a direct, real monetary value.
In other cases, they may deal with contractual issues, again with strong economic implications, as in the case of document certification, supply chain management, voting systems.
%
In most cases, DApps are business critical, and very strict security requirements should be assured.
However, due to the peculiarities of blockchains, DApps development differs from the development of standard software. For instance, once deployed, the code of a SC cannot change. 
Therefore, security plays a major role in the development process of DApps. 
Several failures and attacks occurred in recent years highlight the need of new specific software development techniques addressing DApp security.

We strongly believe in the need of code inspection, security patterns, and tests to get an improved security. 
While various research papers discuss smart contract design patterns~\cite{bart2017, wohrer2018, xu2018pattern} or apply them to specific domains~\cite{zhang2017applying, liu2018applying}, our goal is to provide users with a guide that helps them during the smart contract development process, allowing developers to easily verify if they applied all the relevant security patterns and best practices to their smart contracts.

\paragraph{Contributions}
Starting from general concepts and best practices of SC security, this work mainly aims to provide the reader with security assessment checklists that can be easily used for the development of SCs, in the realm of Ethereum and Solidity.
Our main contributions are the following:
\begin{enumerate} 
\item we discuss the critical issues regarding the safety of DApps and present \mainPatterns main security patterns;
\item we provide a security assurance checklist for the design phase composed of \designBestPractice best practices;
\item we provide a security assurance checklist for the coding phase including \codingBestPractice best practices;
\item we provide a security assurance checklist for the testing phase including \testingBestPractice best practices.
\end{enumerate}
In order to both keep our checklists updated, and provide them as a tool that can be easily used by developers, we also release the checklists in a spreadsheet file available at the following link: \href{http://tiny.cc/security\_checklist}{http://tiny.cc/security\_checklist}. These checklists can be customised, removing the not relevant parts, on a project-by-project basis.

\paragraph{Structure of the paper}
The reminder of this paper is organized as follows. ~\Cref{sec:related} discusses some related works.
~\Cref{sec:general} discusses general concepts and critical security issues of DApps.
~\Cref{sec:security} discussed the security patterns and presents our best practice checklists. 
Specifically,~\Cref{subsec:security_abstract} presents the main security patterns;
~\Cref{subsec:security_design} focuses on the design phase and presents our first assurance checklist;
~\Cref{subsec:security_coding} discusses the coding phase and presents our second assurance checklist.
~\Cref{subsec:security_testing} presents the testing phase and the related checklist.
~\Cref{subsec:security_others} discusses other design patterns that we found in literature.
Finally,~\Cref{sec:conclusions} draws some conclusions.

\section{Related Work}
\label{sec:related}
Various papers describe some security vulnerabilities that affect SCs~\cite{atzei2017, li2020survey}.
Thereafter, there was a growing literature on techniques for improving SC security, such as formal verification~\cite{atzei2019developing, mavridou2018designing, mavridou2019verisolid, harz2018towards}, static analysis~\cite{luu2016making, kalra2018zeus, brent2018vandal}, dynamic analysis~\cite{chen2018understanding}.
Praitheeshan et al. provides a survey on such techniques~\cite{prait2019}.
Our work focuses on software engineering for SCs development and proposes security patterns checklists to be used in the design, coding and testing phases.
Accordingly, we distinguish two main categories of related works, which we describe below.

The first category includes the works on software engineering for SCs development. This area of research is quite recent: Porru et al. highlighted the need of Blockchain-Oriented Software Engineering (BOSE) in 2017~\cite{porru2017}.
In their overview of blockchain-based systems for digital coupons, Podda et al. analyse the source code of SCs and highlight the need of design patterns to help implementing these systems in a safe and effective way~\cite{PoddaPompianu2020}.
Chakraborty et al. confirmed that standard software engineering methods need to be updated when developing blockchain software~\cite{chakraborty2018, bosu2019}.
Various papers propose methods to facilitate the blockchain development process~\cite{jurg2019, Wessling2018, fridgen2018}. 
An overview of engineering aspects of blockchains is introduced in~\cite{xu2017} and extended in~\cite{xu2019}. 

The second category includes the works which study design patterns for smart contracts.
Bartoletti et al. inspect the source code of 811 Solidity SCs and identify nine common design patterns~\cite{bart2017}.
In~\cite{zhang2017applying} Peng et al. present four design patterns and describe how they apply to the blockchain-based healthcare domain.
Wohrer et al. elaborate six security patterns which describe solutions to typical security issues~\cite{wohrer2018}. They also provide the Solidity source code of each pattern. 
Xu et al. show a pattern collection including fifteen patterns and classify them into four categories: interaction between blockchain and real world, data management, security, and contract structural patterns~\cite{xu2018pattern}.
Liu et al. summarise eight design patterns and group them into four categories: creational, structural, inter-behavioral, and intra-behavioral patterns~\cite{liu2018applying}.
%
In ~\cite{gasSaving} Marchesi et al. provide a pattern collection to help gas saving in developing Smart Contracts on Ethereum and classify them into five categories, based on their features.

Compared with the existing surveys~\cite{wohrer2018, xu2018pattern}, our work classifies design patterns according to the different stages of the software development process rather than the type of issue addressed by each pattern. 
Since we include in our analysis the patterns relevant to security issues that we found in literature, several patterns presented in~\cite{zhang2017applying, liu2018applying, bart2017, wohrer2018, xu2018pattern} are the same, or are similar to the patterns we discuss. Furthermore, we notice that sometimes a pattern is presented by different authors with a different name. 
We provide a deep comparison of our patterns with respect to the patterns discussed in the existing works in~\Cref{sec:security}. Specifically, the \textit{Ref.} column of our checklists provides a reference to all the papers in which we identify each pattern.

\section{General concepts of DApp security} \label{sec:general}


Assessing and defining patterns of good programming practice for SCs for granting security in DApps is still in its infancy and is an ongoing area of research. Nevertheless, based on the programmers' experience and on recent exploited weaknesses --very (in)famous and critical also for the amount of real money involved--, some major advices for security assessment in SCs have been identified and discussed among the Solidity developers community.
In fact, Ethereum and Blockchain ecosystem are highly new and still somewhat experimental. 
In addition, SCs are often designed to handle and transfer significant amount of money (in cryptocurrency, but easily exchangeable to real money). 
Therefore, it is necessary that they correctly achieve their purposes, but it is also crucial that their execution is secure against attacks.

The critical issues regarding the safety of a DApp can be divided in three areas:
\begin{itemize}
\item \textit{Issues related to Blockchain itself}: the blockchain itself could be attacked. It is known, for instance, that blockchains using proof-of-work for block generation are subject, at least theoretically, to the so-called "51\% attack". Those based on proof-of-stake are vulnerable to other types of attack, for example to "fake stake attack". Using Ethereum technology, the use of the main net lowers the probability of a "51\% attack", given the number and the computing power fielded by the miners. Instead, using Ethereum Classic blockchain, a fork derived from Ethereum in 2016, the probability is higher because its miners' computer power is much lower. Using a permissioned blockchain, for instance Ethereum Parity "proof-of-authority", the blockchain security depends on the honesty and reliability of the validating members, and on their control over their respective IT services. Clearly, this kind of attacks are more a problem of design choice of the technology to be used than of proper DApp design, so their prevention go beyond the scope of this paper.

\item \textit{Issues related to SCs}: the most critical part of a DApp are the SCs, whose bytecode is publicly available, and exposed to all possible exploits. Moreover, developers often lack a full knowledge about implementation and usage of SCs, due to the the fact that this technology is in its early stage, it is evolving fast and is different from traditional development. In literature there are several analyses of possible vulnerabilities related to both Ethereum virtual machine and Solidity language~\cite{prait2019,huang2019,liu2019}. These are a good starting point for providing a checklist of patterns to verify the SCs under development.

\item \textit{Issues related to the App System}: The App System is composed of the server and client side of the DApp, interacting with the SCs on one side, and with human actors, IoT devices and other systems on the other side. It must be designed and implemented with care, but it is somewhat less critical, provided that all best practices related to the security of Web applications are used; a special emphasis must be made to safeguard the access to the private keys of the various actors. 
\end{itemize}

In this paper we tackles the issues related to SCs development which are 
the most critical, common and interesting for blockchain software developers. 

The first and foremost concept in security management is to have a security mindset. 
The development team(s), and the whole organization, must be fully aware of the importance of security and protection from attacks.
A good starting point to focus on security are the Top 10 Proactive Controls of OWASP organization~\cite{owasp2018}. 
Among them we identified those most relevant for DApp security, often neglected in SCs development, and report them ordered by importance:
\begin{itemize}
\item C1: \textbf{Define Security Requirements.} This looks straightforward, but it is often underestimated. You must explicitly define the security requirements needed for your system. The requirements can be written as User Stories, or as non-functional features, and should have acceptance tests in the form of test cases to confirm these requirements have been implemented.
\item C2: \textbf{Leverage Security Frameworks and Libraries.} Don't write everything from scratch, but reuse software that is security-hardened, is coming from trusted sources and is maintained up to date.
\item C5: \textbf{Validate All Inputs.} This should be performed for user inputs on server-side, because client-side validation can be bypassed. Also, let the SC itself perform validation of key data sent to it through messages.
\item C6: \textbf{Implement Digital Identity.} In a DApp environment, digital identities are guaranteed by addresses and by the ownership of the relative private key, so this control is quite straightforward. Nevertheless specific checks 
for address ownership must be implemented to grant SCs security
from unauthorized uses. 
\item C7: \textbf{Enforce Access Controls.} SC can check access levels of addresses through a mapping, and act accordingly.
\item C8: \textbf{Protect Data Everywhere.} In particular, be aware that data stored in a SC are always accessible to read, independently of their visibility.
\item C10: \textbf{Handle All Errors and Exceptions.} It is known that even small mistakes in error handling, or forgetting to handle errors can lead to catastrophic failures in distributed systems. This is particularly true for SCs.
\end{itemize}

The general guidelines reported by the Solidity best practices~\cite{bestPractices}, section "General Philosophy", which complement OWASP ones, are specifically related to SC security: 
\begin{enumerate}
\item \textbf{Prepare for failure.} Be able to respond to errors, also in the context of SCs, which cannot be changed once deployed. 
\item \textbf{Rollout carefully.} Try your best to catch and fix the bugs before the SC is fully released. Test contracts thoroughly, and add tests whenever new attack vectors are discovered.
\item \textbf{Keep SCs simple.} Complexity increases the risk of errors, so ensure that SCs and functions are small and modular, reuse SCs that are proven, prefer clarity to performance.
\item \textbf{Keep up to date.} Keep track of new security developments and upgrade to the latest version of any tool or library as quickly as possible.
It can be hard to enforce this security pattern once the SC has been deployed, since the code 
cannot be directly updated. Thus, special care 
must be devoted to this pattern before deploy. 
\item \textbf{Be aware of blockchain properties.} While your previous programming experience is also applicable to SC programming, there are several pitfalls to be aware of. An example can be the
well known Parity Wallet ``Hack'' occurred in November 2017, where the (apparently incidental) use of a self-destruct function by a user who took ownership of the library frozen all Parity multisig wallets and all the cryptocurrency kept in there. 
\end{enumerate}

Although the literature provides various lists of guidelines and best practices to improve security, it is not trivial to apply them to the SC development process. For this reason, our focus is to provide users with checklists that can be easily adopted during the development activity.
In the following of this paper, we focus on security assurance practices regarding SC design, coding, and testing. Indeed, issues related to SCs are the most critical and less studied among the three categories of issues cited above.
Specifically, performing our analysis, we took into account both the Proactive Controls of OWASP and the general guidelines reported above.

\section{Security assessment for Smart Contracts} 
\label{sec:security}
In this section we present the security pattern collection and the three security checklists to be performed during the different phases of the development process. 
First, \Cref{subsec:description} and \Cref{subsec:security_collecting} present our methodology for collecting patterns. \Cref{subsec:security_abstract} discusses the abstract security patterns. The remainder of the section presents the three checklists for the design (\Cref{subsec:security_design}), coding (\Cref{subsec:security_coding}), and test (\Cref{subsec:security_testing}) phases.
~\Cref{subsec:security_others} presents other design patterns discussed in literature.

\subsection{Security: patterns and best practices}
\label{subsec:description}

Our aim is to provide the reader with an easy way to verify that all the known security issues are managed. 
Several related works provide various practices to mitigate these issues. 
They often refer to all these practices as \textit{patterns}. 
Indeed, most of these items are very simple and not structured to be properly called patterns.

Accordingly, we divide them into two different categories: \textit{abstract security patterns} and \textit{best practices}.
The first category includes patterns that are not related to a specific step of the development process.
The second category includes those items related to a specific phase of the smart contract development process.
Moreover, from the best practices, it is easy to extract checklists to be used to perform security assurance during each development phase. 
We provide the complete checklists online at \href{http://tiny.cc/security\_checklist}{http://tiny.cc/security\_checklist}. 
These checklists can be customized based on the projects requirements.
These practices should complement a broader software development process for producing DApps, such as an agile one.

~\Cref{tab:secPat} collects the abstract security patterns. For each pattern, we provide a brief description of both the problem addressed and its solution, along with a list of references to the papers discussing it. A short discussion is also presented in Sec. \ref{subsec:security_abstract}.
Tables~\ref{tab:design},~\ref{tab:coding}, and~\ref{tab:testing} present the security assurance practices we propose. They describe the checks to be performed (column \textit{Name}), a short description of the vulnerabilities and how to avoid them, one or more references to learn more about the problem and a reference to the related security patterns shown in ~\Cref{tab:secPat}. 

\subsection{Collecting design patterns}
\label{subsec:security_collecting}

Our methodology for collecting design patterns and building our best practices checklists is described in the following steps:

\begin{enumerate}
    \item We queried on June 1st, 2020 the Web, searching for design patterns and best practices for smart contracts.
    
    \item We then manually inspected the results, collecting several data sources gathering different lists of patterns. Most of the sources are scientific papers, but we also found some forum articles.
    
    \item We built a first list of \totItems items potentially relevant to our work, by merging all the lists collected in the previous step.
    
    \item We filtered out our list, by excluding \otherPatterns items that, although are useful patterns, are not directly related to security aspects. For the sake of completeness, we decide to include those items in~\Cref{tab:other} and discuss them in~\Cref{subsec:security_others}.
    
    \item We analysed all the remaining items and splitted them into two main categories: \mainPatterns \textit{abstract security patterns} and \totBestPractice \textit{best practices}.
    
    \item We splitted the best practices into three groups according to the related phase in the development process: design, coding, testing and deployment.
    
\end{enumerate}

Summing up, we identified \mainPatterns abstract security patterns, and \totBestPractice best security practices: \designBestPractice for designing, \codingBestPractice for coding, and \testingBestPractice for testing and deployment of SCs. 
The other design patterns not related to security are \otherPatterns.

\subsection{Abstract security patterns}
\label{subsec:security_abstract}

Patterns are schemes of a standard solution to a recurring problem, with a certain structure. However, the security patterns in the DApps scenario, do not always have a structure and uniformity comparable to traditional design patterns~\cite{gamma1994}. 
Some of the patterns we present are variants of well-know traditional design patterns, applied in the blockchain and smart contracts scenario, whereas others are specifically designed to address the peculiarities of Ethereum and Solidity context.

We describe below some of the patterns we deem to be the most significant.
A complete description of all the patterns is available in~\Cref{tab:secPat}.

An example of traditional design pattern which is useful also in the blockchain context is the \textit{Proxy} pattern. 
It introduces the possibility of upgrading a contract, which is by nature immutable, without breaking any dependency. 
The idea is to divide the contract into modules and to use a Proxy contract to delegate calls to specific modules. In this way, indeed, it is possible to modify a contract by implementing a new version and replacing its address with the previous one in the archive stored in the Proxy.

An example of pattern specific of the blockchain context is the \textit{Oracle} pattern, which is introduced to respond to the need to acquire information from the outside world. 
In fact, a smart contract is by nature isolated, and cannot acquire information directly.
This is due to the fact that network nodes must agree on the state of transactions. To accomplish this, nodes should evaluate only \textit{static} data. 
On the contrary, the outside world, for instance a website API, could provide different responses to the same query performed by different network nodes, breaking blockchain consensus. 
An oracle is a smart contract handled by a trusted authority for uploading outside data to the blockchain. %
This allows network nodes to query the static blockchain data instead of the outside world, overcoming the limits described above.

Another  pattern particularly important in a decentralised context is the \textit{Authorization} pattern. 
Since SCs are publicly accessible to all blockchain participants, it is critical to restrict authorizations to perform specific tasks. 
Specifically, for each contract method, developers must specify the subset of participants who can call it.
Contracts usually define at least one \textit{contract owner}, which is the only entity authorised to call critical methods. 
~\Cref{tab:coding} shows various techniques to handle authorizations properly during the coding phase.

A \textit{Time constraint} defines when the related action is allowed to be performed. Different blockchain nodes could process transactions at different timestamps, due to network latency or further causes. Consequently, part of the network could consider an action as executed beyond the time constraint, and the corresponding transaction could be rejected by the whole network.
Accordingly, developers must set time constraints carefully, for instance by ensuring that there is enough time between two consecutive time constraints.

\subsection{Security in the design phase}
\label{subsec:security_design}
In the design phase, developers must be aware of, and use security patterns, as reported in references~\cite{wohrer2018,bart2017,proxyy2019}, which we refer to. 
In the remainder of this subsection we describe some of the design best practices. For the sake of brevity we describe a few of them, while~\Cref{tab:design} shows the full list.

A \textit{fail-safe mechanism} is a function that allows contract owners to disable specific SC methods.
Developers should always design mechanisms to either update or terminate contracts because, due to the immutability of blockchains, SCs can not be removed once published. 
The fail-safe best practice can be exploited to accomplish several security patterns. 
For instance, it can be used for terminating a contract (\textit{Termination} pattern), and optionally for enabling a new version of the contract (\textit{Proxy delegate}).
Moreover this mechanism could be used for slowing down sensitive tasks (\textit{Speed Bump}, \textit{Rate Limit}).
Usually, this mechanism is enabled by the contract owner (\textit{Ownership}).



\subsection{Security in the coding phase}
\label{subsec:security_coding}
Below, we describe some of the most representative coding best practices.
The full checklist for security assessment in the coding phase is reported in~\Cref{tab:coding}.

During coding, one major class of problems derives from \textit{external calls}, namely from functions which recur to others' SC code for completing their execution. 
In fact, a SC can call another SC, exploiting the execution of code contained in the latter contract. The pattern can be recursive, so the called SC can in turn perform an external call, and so on. 
As a consequence, external calls must be treated like calls to `untrusted' software. 
They should be avoided or minimized, because some malicious code could be introduced somewhere in a SC belonging to this path, and any external call represents a security risk. 

A typical risk of such contract interaction is \textit{reentrancy}, namely the called contract can call back the calling function before the overall function execution has been completed. This pattern has been performed in the DAO attack. 
When it is not possible to avoid external calls, label all the potentially unsafe variables, functions and contracts interfaces as "untrusted". 

In order to prevent these issues, check accurately all preconditions and restrict concurrent accesses to resources, as defined by the \textit{Check-effect-interaction} and \textit{Mutex} patterns.

Another important best practice for SC security and error handling is to \textit{validate inputs} by using \textit{assert()}, \textit{require()} and \textit{revert()} guard functions. 
They are a very powerful security tool, and are the subject of security pattern \textit{Guard Check} presented in~\Cref{tab:secPat}.
In general, use \textit{assert()} to check for invariants, to validate state after making changes, to prevent wrong conditions; if an \textit{assert()} statement fails, something very wrong happened and you need to fix the code.
Use \textit{require()} when you want to validate: user inputs, state conditions preceding an execution, or the response of an external call.
Use \textit{revert()} to handle the same type of cases as \textit{require()}, but with more complex logic~\cite{bestPractices}.

\subsection{Security in the testing and deployment phases}
\label{subsec:security_testing}
In this subsection we focus on the testing and deployment steps, and describe some of the best practices shown in~\Cref{tab:testing}.

As in traditional software engineering, also in the smart contract context \textit{develop unit testing} is important.
Currently, there are several techniques for testing smart contracts.
We do not prescribe the use of specific testing practices, such as Test Driven Design, but we highlight the importance of testing.
One way is to use a browser-based real-time compiler and runtime environment for Solidity, such as Remix.

Another way is to \textit{use frameworks for testing}.
Presently, among the most popular testing frameworks for Ethereum DApps there are Truffle ~\cite{tru2019}, Embark ~\cite{embark} and Etherlime ~\cite{etherlime}. 
Moreover, the Ethereum community operates multiple \textit{test networks}. 
These are used by developers to test applications under different conditions before deploying them on the main Ethereum network. 
The most famous ones are Ropsten ~\cite{ropsten} , Rinkeby ~\cite{rinkeby} and Goerli ~\cite{goerli}.
Finally, it is possible to set up a local Ethereum blockchain which can be used to run tests, execute commands, and inspect state while controlling how the chain operates. 
An example of such a software is Ganache ~\cite{ganache}, from the Truffle suite.

\subsection{Other patterns}
\label{subsec:security_others}
As described in our methodology (\Cref{subsec:security_collecting}), \otherPatterns more patterns emerged from the literature. Those patterns are not strictly related to contract security but are useful for a good SC design.
Therefore, we decided to include them in this work for completeness and briefly describe some of them in the followings. \Cref{tab:other} shows the full list of those patterns.

The \textit{Publisher-Subscriber} pattern is a well-known design pattern studied in traditional (i.e. not blockchain-based) software engineering. 
According to this pattern, when a module needs to receive some messages from other software modules (e.g. a smart contract), developers should implement a messaging infrastructure that allows each module to be easily notified when a new message is generated.

The \textit{Tokenisation} pattern suggests to use tokens for representing good and services in the blockchain. 
The Ethereum ecosystem provides standards to handle several types of tokens, such as the ERC20 and ERC721.

\section{Conclusions}
\label{sec:conclusions}
In this paper we presented a pattern collection of \totUsefulItems patterns and best practices for security in blockchain-based applications.
We focused on Ethereum and Solidity, since they are nowadays the most used blockchain and programming language to develop DApps.
Our collection includes \mainPatterns abstract patterns and \totBestPractice best practices: \designBestPractice for the design, \codingBestPractice for the coding, and \testingBestPractice for the testing phases of smart contract development process.
Some patterns are adaptations of traditional software patterns modified to account for SCs specificity in the context of Solidity language,
others are specifically designed for DApps, considering the unique properties of the blockchain.

To provide a useful guide that allows developers to easily verify if they applied all the relevant security patterns to their projects, we collected the patterns into three security assurance checklists, based on the phase.
These checklists can be customised on a project-by-project basis.
In order to keep them updated and to provide an easy to use tool, we also released these checklists in a spreadsheet file available at the following link: \href{http://tiny.cc/security\_checklist}{http://tiny.cc/security\_checklist}. 
For completeness, since our collection excluded \otherPatterns of the patterns we found in the literature, we decided to present them.

Combining good software design practice with security assurance checklists, developers can create DApps that are more reliable, address security issues and mitigate typical attacks.


Future works will extend these security checklists in order to address other blockchains, such as Hyperledger Fabric and other permissioned blockchains.
Indeed, permissioned systems introduce new challenges that affect the whole development process, such as the designing and the maintainability of the network.
Another research direction is to gather feedback by developers and provide statistics to improve the proposed checklists.

Finally, we will keep updated the proposed checklists at our link: \href{http://tiny.cc/security\_checklist}{http://tiny.cc/security\_checklist}.

\begin{center}
\onecolumn
\topcaption{Abstract security patterns} \label{tab:secPat}
\tablefirsthead{
\hline
\multicolumn{1}{|c|}{\textbf{ID}} & 
\multicolumn{1}{c|}{\textbf{Name}} & \multicolumn{1}{c|}{\textbf{Description}} & \multicolumn{1}{c|}{\textbf{Ref.}} \\ 
\hline}
\begin{supertabular}{|p{1cm}|p{3cm}|p{10cm}|p{1cm}|}
CEI & \textit{Check Effect Interaction} & When performing a function in a SC: first, check all the preconditions, then apply the effects to the contract's state, and finally interact with other contracts. Never alter this sequence. &~\cite{wohrer2018} \\  \hline

PD & \textit{Proxy Delegate} / \textit{Decorator}
& Proxy patterns are a set of SCs working together to facilitate upgrading of SCs, despite their intrinsic immutability. A Proxy is used to refer to another SC, whose address can be changed. This approach also ensures that blockchain resources are used sparingly, thus saving GAS. &~\cite{proxyy2019, ethsec2019, zhang2017applying, liu2018applying, gasSaving} \\  \hline

AU & \textit{Authorization} & Restrict the execution of critical methods to specific users. This is accomplished using mappings of addresses, and is typically checked using modifiers. &~\cite{bart2017} \\  \hline

OW & \textit{Ownership} & Specify the contract owner, which is responsible for contract management and has special permissions, e.g. it is the only address authorized to call critical methods. This patter can be seen as a special instance of the authorization pattern. &~\cite{bart2017} \\  \hline

OR & \textit{Oracle} & An oracle is a SC providing data from outside the blockchain, which are in turn fed to the oracle by a trusted source. Here the security risk lies in how actually the source can be trusted. &~\cite{bart2017, xu2018pattern} \\  \hline

RO & \textit{Reverse Oracle} & A reverse oracle is a SC providing data to be read by off-chain components for checking specific conditions. &~\cite{xu2018pattern} \\ \hline

RL & \textit{Rate Limit} & Regulate how often a task can be executed within a period of time, to limit the number of messages sent to a SC, and thus its computational load. &~\cite{wohrer2018} \\  \hline

BL & \textit{Balance Limit} & Limit the maximum amount of funds held within a SC. &~\cite{wohrer2018} \\  \hline

GC & \textit{Guard Check} & Ensure that all requirements on a SC state and on function inputs are met. &~\cite{ethsec2019} \\ \hline

TC & \textit{Time Constraint} & A time constraint specifies when an action is permitted, depending on the time registered in the block holding the transaction. It could be also used in Speed Bump and Rate Limit patterns. &~\cite{bart2017} \\  \hline

TE & \textit{Termination} & Used when the life of a SC has come to an end. This can be done by inserting ad-hoc code in the contract, or calling the selfdestruct function. Usually, only the contract owner is authorized to terminate a contract. &~\cite{bart2017} \\  \hline

MH & \textit{Math} & A logic which computes some critical operations, protecting from overflows, underflows or other undesired characteristics of finite arithmetic. &~\cite{bart2017} \\  \hline

PR & \textit{Privacy} & Encrypt on-chain critical data improving confidentiality and meeting legal requirements, such as the European GDPR. &~\cite{xu2018pattern} \\ \hline

REU & \textit{Reusability} & Use contract libraries and templates as a factory for creating multiple instances.  &~\cite{xu2018pattern} \\ \hline

MU & \textit{Mutex} & A mutex is a mechanism to restrict concurrent access to a resource. Utilize it to hinder an external call from re-entering its caller function again. &~\cite{wohrer2018} \\  \hline

SB & \textit{Speed Bump} & Slow down contract sensitive tasks, so when malicious actions occur, the damage is limited and more time to counteract is available. For instance, limit the amount of money a user can withdraw per day, or impose a delay before withdrawals. &~\cite{wohrer2018} \\  \hline

\end{supertabular}
\end{center}

\begin{center}
\onecolumn
\topcaption{Security assurance checklist for the design phase} \label{tab:design}
\tablefirsthead{
\hline
\multicolumn{1}{|c|}{\textbf{Name}} & \multicolumn{1}{c|}{\textbf{Description}} & \multicolumn{1}{c|}{\textbf{Ref.}} & \multicolumn{1}{c|}{\textbf{Related patterns}} \\ 
\hline}
\begin{supertabular}{|p{2.5cm}|p{10cm}|p{2cm}|p{2.5cm}|}
\hline

\textit{Include fail-safe mechanisms} & It is important to have some way to update the contract in the case some bugs will be discovered. Incorporate an emergency stop functionality into the SC that can be triggered by an authenticated party to disable sensitive functions. The fail-safe mechanism, if implemented using the \textit{Proxy Delegate}, could be also exploited for forwarding calls and data to another contract, which is an updated version of the current one (for instance, a version where the bug has been fixed). &~\cite{wohrer2018} & SB, RL, TE, PD, OW \\  \hline

\textit{Never assume that a contract has zero balance} & Be aware of coding an invariant that strictly checks the balance of a contract. An attacker can forcibly send ether to any account and this cannot be prevented. &~\cite{bestPractices} & CEI, MH, GC \\  \hline  

\textit{State Channel} / \textit{Off-chain Support} & In some contexts, transactions either have too high fee compared to their value, or must have low latency. In these cases, rather than performing each blockchain transaction, it is possible to firstly perform the operations outside the blockchain, and then register all the results batching the requests in a unique blockchain transaction. &~\cite{xu2018pattern, PoddaPompianu2020} & RL \\ \hline

\textit{Limit the amount of ether} & If the code, the compiler or the platform has a bug, the funds stored in your smart contract may be lost, so limit the maximum amount. Check that all money transfers are performed through explicit withdrawals made by the beneficiary. & ~\cite{atzei2017} & RL, BL, AU \\  \hline

\textit{Beware of transaction ordering} & Miners have the power to alter the order of transactions arriving in short times. Inconsistent transactions' orders, with respect to the time of invocations, can cause race conditions.&~\cite{prait2019} & TC \\ \hline

\textit{Be careful with multiple inheritance} & Solidity uses the "C3 linearization". This means that when a contract is deployed, the compiler will linearize the inheritance from right to left. Multiple overrides of a function in complex inheritance hierarchies could potentially interact in tricky ways. &~\cite{bestPractices} & PD, REU \\ \hline

\textit{Use trustworthy dependencies} & Use audited and trustworthy dependencies to existing SCs and ensure that newly written code is minimized by using libraries. & ~\cite{bestPractices} & REU \\ \hline

\textit{Withdrawal from Contracts} / \textit{Pull over Push} & When you need to send Ethers or tokens to an address, don't send them directly. Instead, authorize the address' owner to withdraw the funds, and let s/he perform the job. &~\cite{solidity2019, ethsec2019} & CEI \\ \hline

\end{supertabular}
\end{center}

\begin{center}
\onecolumn
\topcaption{Security assurance checklist for the coding phase} \label{tab:coding}
\tablefirsthead{
\hline
\multicolumn{1}{|c|}{\textbf{Name}} & \multicolumn{1}{c|}{\textbf{Description}} & \multicolumn{1}{c|}{\textbf{Ref.}} & \multicolumn{1}{c|}{\textbf{Related patterns}} \\ 
\hline}
\begin{supertabular}{|p{2cm}|p{10cm}|p{1cm}|p{1cm}|}
\hline

\textit{Be careful with external calls} & If possible, avoid them. When using low-level call functions
make sure to handle the possibility that the call will fail, by checking the return value. Also, avoid combining multiple ether transfers in a single transaction. Mark untrusted interactions: name the variables, methods, and contract interfaces of the functions that call external contracts, in a way that makes it clear that interacting with them is potentially unsafe.
&~\cite{bestPractices} & CEI, MU, GC \\ \hline

\textit{Beware of re-entrancy} & Never write functions that could be called recursively, before the first invocations is finished. This may cause destructive consequences. Ensure state committed before an external call. &~\cite{atzei2017, bestPractices} & CEI, MU \\ \hline

\textit{Embed addresses to grant permissions} & Make sure that critical methods can be invoked only by a specific set of addresses, which belong to privileged users. For instance, each contract has an owner and only this address can invoke certain methods, like the method for updating the address of the owner of the contract. &~\cite{xu2018pattern} & AU, OW \\ \hline

\textit{Use hash secrets to grant permissions} & Sometimes you need to provide authorizations to some authorities whose addresses are not known yet in the developing phase (for instance, they are unknown authorities). Although the \textit{Embed permissions} pattern can not be applied, hash secrets help providing user permissions without specifying any address. First, generate a secret key and, in the contract, provide permissions by requiring its hash. Then, send (off-chain) the secret key to the authorities you want to grant permissions.  &~\cite{liu2018applying, xu2018pattern} & AU \\ \hline

\textit{Use multi-signature} & Define a set of entities (or addresses) that can authorize an action and require that only a subset of them is required to authorize the action. &~\cite{liu2018applying, xu2018pattern} & AU, OW \\ \hline

\textit{Avoid using tx.origin for authorizations} & \textit{tx.origin} is a global variable that returns the address of the message sender. Do not use it as an authorization mechanism. &~\cite{Mense:2018:SVE:3282373.3282419} & AU \\  \hline

\textit{Encrypt on-chain data} & Encrypt blockchain data for improving confidentiality and privacy. This is particularly important when actors are in competition. & \cite{xu2018pattern} & PR \\ \hline

\textit{Hash objects for tracking off-chain data} & Large objects (such as videos) should not be embedded in the blockchain, their hashes can be easily uploaded instead. Hashing objects can be also applied to hide sensitive data in order to meet specific legal requirements, such as the European GDPR. & \cite{xu2018pattern} & PR \\ \hline

\textit{Use platform related standards} & Use platform related standards, like the ERC (Ethereum Request for Comment) standards, which are application-level blueprints and conventions in the Ethereum ecosystem. & \cite{xu2018pattern} & REU \\ \hline

\textit{Prevent overflow and underflow} & If a balance reaches the maximum uint value it will circle back to zero; similarly, if a uint is made to be less than zero, it will cause an underflow and get set to its maximum value. One simple solution to this issue is to use a library like \textit{SafeMath.sol} by OpenZeppelin. &~\cite{prait2019} & MH, GC, REU, BL \\ \hline

\textit{Beware of rounding errors} & All integer divisions round down to the nearest integer. Check that truncation does not produce unexpected behavior (locked funds, incorrect results). &~\cite{bestPractices} & MH, GC, REU \\ \hline

\textit{Validate inputs to external and public functions} & Make sure the requirements are verified and check for arguments. Use properly \textit{assert()}, \textit{require()} and \textit{revert()} to check user inputs, SC state, invariants. &~\cite{prait2019, ethsec2019} &  GC \\  
\hline
\textit{Prevent unbounded loops} & When executing loops, the gas consumed increases with each iteration until it hits the block's gasLimit, stopping the execution. Accordingly, plan the number of iterations you need to perform and establish a maximum number. If you still need more iterations, divide computation among distinct transactions. &~\cite{ethsec2019, gasSaving} & RL, BL, TC, TE \\ \hline

\textit{Provide fallback functions} & The "fallback function" is called whenever a contract receive a message which does not match any of the available functions, or whenever it receives Ethers without any other data associated with the transaction.  Remember to mark it as \textit{payable}, be sure it does not have any arguments, has external visibility and does not return anything. Moreover, keep it simple and if the fallback function is intended to be used only for the purpose of logging received Ether, check that the data is empty (i.e. require(msg.data.length == 0) ).
&~\cite{bestPractices, prait2019} & CEI, MU, GC \\ \hline

\textit{Check if built-in variables or functions were overridden} & It is possible to override built-in globals in Solidity. This allows SCs to override the functionality of built-ins such as \textit{msg} and \textit{revert()}. Although this is intended, it can mislead users of a SC, so the whole code of every SC called from the SC you are writing must be checked. &~\cite{bestPractices} & GC \\ \hline

\textit{Use interface type instead of the address for type safety} & When a function takes a contract address as an argument, it is better to pass an interface or contract type rather than a raw \textit{address}. If the function is called elsewhere within the source code, the compiler will provide additional type safety guarantees. &~\cite{bestPractices} & GC \\ \hline

\textit{Be careful with randomness} & 
Random number generation in a deterministic system is very difficult. Do not rely on pseudo-randomness for important mechanisms. Current best solutions include hash-commit-reveal schemes (ie. one party generates a number, publishes its hash to "commit" to the value, and then reveals the value later), querying oracles, and RANDAO. &~\cite{bart2017, huang2019} & OR, REU \\ \hline

\textit{Be careful with Timestamp} & Be aware that the timestamp of a block can be manipulated by a miner; all direct and indirect uses of timestamp should be analyzed and verified. If the scale of your time-dependent event can vary by 30 seconds and maintain integrity, it is safe to use a timestamp. This includes thing like ending of auctions, registration periods, etc. Do not use the \textit{block.number} property as a timestamp. &~\cite{prait2019} & TC \\  \hline

\end{supertabular}
\end{center}

\begin{center}
\onecolumn
\topcaption{Security assurance checklist for the testing and deployment phases} \label{tab:testing}
\tablefirsthead{
\hline
\multicolumn{1}{|c|}{\textbf{Name}} & \multicolumn{1}{c|}{\textbf{Description}} & \multicolumn{1}{c|}{\textbf{Ref.}} \\ 
\hline}
\begin{supertabular}{|p{2.5cm}|p{10cm}|p{2cm}|}
\textit{Fix compiler warnings} & Take warnings seriously and fix them. Always use the latest version of the compiler to be notified about all recently introduced warnings. &~\cite{solidity2019} \\  \hline

\textit{Lock programs to specific compiler version} & Contracts should be deployed with the same compiler version and flags that they have been tested with, so locking the version helps avoid the risk of undiscovered bugs. &~\cite{bestPractices} \\ \hline

\textit{Enforce invariants with assert} & An assert guard triggers when an assertion fails - for instance an invariant property changing. You can verify it with a call to \textit{assert()}. Assert guards should be combined with other techniques, such as pausing the contract and allowing upgrades. (Otherwise, you may end up stuck, with an assertion that is always failing.) &~\cite{bestPractices} \\  \hline

\textit{Develop unit testing} & Be sure to have a $100\%$ text coverage and cover all critical edge cases with unit tests. Do not deploy recently written code, especially if it was written under tight deadline. &~\cite{bestPractices} \\  \hline

\textit{Use frameworks for testing} & When approaching smart contract testing, do not start from scratch but use existing framework for contract testing. &~\cite{bestPractices} \\ \hline

\textit{Use test networks} & Before deploying the smart contract in the main network, try it in a public test network or use a software for configuring a private local network. &~\cite{bestPractices} \\ \hline

\end{supertabular}
\end{center}

\begin{center}
\onecolumn
\topcaption{Other design patterns} \label{tab:other}
\tablefirsthead{
\hline
\multicolumn{1}{|c|}{\textbf{Name}} & \multicolumn{1}{c|}{\textbf{Description}} & \multicolumn{1}{c|}{\textbf{Ref.}} \\ 
\hline}
\begin{supertabular}{|p{2.5cm}|p{10cm}|p{2cm}|}
\textit{Publisher-Subscriber} & When a state change must trigger a computation in a different object, implement a messaging infrastructure where the contracts that produce messages (called \textit{publishers}) can generate messages and the other contracts (called \textit{subscribers}) receive them. The pattern, also known as \textit{Observer}, reduces the overhead of constant information filtering. &~\cite{zhang2017applying} \\ \hline

\textit{Tokenisation} & Use tokens for transferring digital or physical services. Use standards, such as ERC20 and ERC721. &~\cite{xu2018pattern} \\ \hline

\textit{X-confirmation} & In order to ensure that a transaction is confirmed (i.e. there is a low probability that a fork happens), wait for new blocks to be added to the blockchain. The amount of blocks depends on the adopted blockchain. &~\cite{xu2018pattern} \\ \hline

\textit{Contract Registry} & Often a smart contract needs to interact with other contracts which can be updated over time. A contract registry maps each smart contract to the address of its latest version. Accordingly, when invoking a smart contract, the correct address should be retrieved from the registry. &~\cite{xu2018pattern} \\ \hline

\textit{Eternal storage} / \textit{Data Contract} & Contract data and logic should be stored into separate contracts. In this way, when the logic need to be updated (by using a new smart contract), there is no need to migrate old data. &~\cite{xu2018pattern} \\ \hline

\textit{Abstract factory} & Sometimes, systems need to work with groups of related contracts, for instance with contracts which represent various level user account. In order to keep the system independent from the different contracts, define an abstract contract for creating all the related contracts. &~\cite{zhang2017applying, xu2018pattern, liu2018applying}
\\ \hline 

\end{supertabular}
\end{center}
\twocolumn

\bibliographystyle{acm}
\bibliography{bibliography.bib}

\end{document}